\documentclass[conference]{IEEEtran}
\usepackage[utf8]{inputenc}
\usepackage[nocompress]{cite}
\usepackage[pdftex]{graphicx}
\usepackage{url}
\usepackage{flushend}
\usepackage{multirow, multicol}
\usepackage{xcolor,colortbl}

\usepackage{amsmath}
\usepackage{algorithmic}
\usepackage{array}
\usepackage{dblfloatfix}
\usepackage[ruled,vlined]{algorithm2e}

\usepackage{soul}
\usepackage{color}

\usepackage{booktabs}
\usepackage{tabularx}

\usepackage{microtype}

\usepackage{nicefrac}

\begin{document}

\title{Dependability Evaluation of Middleware Technology for Large-scale Distributed Caching}

\author{\IEEEauthorblockN{Domenico Cotroneo, Roberto Natella and Stefano Rosiello}
\IEEEauthorblockA{Università degli Studi di Napoli ``Federico II'', Italy\\
\{cotroneo, roberto.natella, stefano.rosiello\}@unina.it}}

%\IEEEspecialpapernotice{Submission type: Practical Experience Report (10 pages)}

\maketitle

\begin{abstract}
Distributed caching systems (e.g., Memcached) are widely used by service providers to satisfy accesses by millions of concurrent clients. Given their large-scale, modern distributed systems rely on a \emph{middleware} layer to manage caching nodes, to make applications easier to develop, and to apply load balancing and replication strategies. 
In this work, we performed a dependability evaluation of three popular middleware platforms, namely \emph{Twemproxy} by Twitter, \emph{Mcrouter} by Facebook, and \emph{Dynomite} by Netflix, to assess availability and performance under faults, including failures of Memcached nodes and congestion due to unbalanced workloads and network link bandwidth bottlenecks. 
We point out the different availability and performance trade-offs achieved by the three platform, and scenarios in which few faulty components cause cascading failures of the whole distributed system.

\end{abstract}

\section{Introduction}
\label{sec:introduction}
% !TEX root = ../main.tex
% !TEX encoding = UTF-8 Unicode

The massive diffusion of social networks, streaming services, and other distributed applications posed new challenges for service providers. Companies such as Facebook, Twitter, and Netflix serve millions of users from all over the world and manage huge amounts of data, achieving at the same time a high level of Quality of Experience (QoE). At the \emph{Velocity} international event \cite{souders2019velocity}, several companies reported  that they face a strict correlation between QoE and users' loss. For instance, Google found that a $400$ ms delay resulted in a $-0.59\%$ change in searches/user. Furthermore, even after the delay was removed, these users still had $-0.21\%$ fewer searches, indicating that bad QoE affects customer behavior in the long term.

To face with the above issue, modern distributed applications adopt a distributed in-memory caching tier between the web/business logic and the storage tiers. Distributed caching offers significant performance benefits at an affordable cost, as main memory becomes cheaper and network interfaces become faster, with 1 Gbit now standard everywhere and 10 Gbit gaining traction. Distributed caches work well on low-cost machines (e.g., machines usually adopted for web servers), as opposed to database servers that require expensive hardware. 
From the software point of view, distributed caches are mostly implemented with key-value datastores, such as \emph{Memcached} \cite{fitzpatrick2004distributed}. These datastores adopt the NoSQL paradigm, where data are stored, retrieved, and managed as associative arrays (also known as ``dictionaries'' and ``hash tables''). They offer several advantages, including flexible data modeling, high performance, and massive scalability.

As distributed systems grow towards millions of components, dealing with failures becomes the standard mode of operation, since there will always be a small but significant number of servers and network segments that are failing at any given time. Therefore, \emph{middleware} platforms \cite{anwar2020customizable} are adopted to enhance large-scale distributed caches with replication and load balancing mechanisms, in order to undertake the burden of tolerating faults and prevent poor QoE experienced by users.

In this work, we investigate the dependability, and in particular performance and fault-tolerance aspects, of three state-of-the-practice \emph{middleware} platforms for distributed caching, namely \emph{Twemproxy} \cite{twemproxyrepo}, \emph{Mcrouter} \cite{mcrouterrepo}, and \emph{Dynomite} \cite{dynomiterepo}, developed by major service providers (respectively Twitter, Facebook, and Netflix).  To the best of our knowledge, previous work has been assessing performance and dependability of NoSQL datastores \cite{beyer2011testing, nelubin2013nosql, ventura2016experimental}, but did not assess these properties in the context of middleware technology for these datastores.

We conducted a large experimental campaign by simulating component failures through fault injection. In addition to the traditional crash failures, we consider the so-called \emph{gray} failures, as reported by several previous studies on incidents in real-world cloud platforms \cite{huang2017gray,gunawi2018fail}. Examples of gray failures are unbalanced overloads and network bottlenecks, which represent subtle failures that are difficult to detect and to mitigate, and that are sometimes overlooked by designers of distributed systems. Our experimental results show that the middleware platforms react differently to the injected faults, as their architectures embrace different trade-offs between performance and availability. These trade-offs are often unclear to designers, due to the lack of experimental evaluations in the literature, which this work aims to compensate for. Our experiments also show that faults of only few components in the datastore tier may cause cascading effects that degrade the performance of the whole distributed system. This result points out that designers of distributed systems need to pay special attention to configure and assess the middleware layer to mitigate these failures.

In the following, Section~\ref{sec:fault_tolerance} provides background on the fault-tolerant middleware platforms; Section~\ref{sec:experiment} presents the experimental methodology and setup; Section~\ref{sec:discussion} discusses the experimental results; Section~\ref{sec:related} discusses related work; Section~\ref{sec:conclusion} concludes the paper.

\section{Middleware platforms for distributed caching}
\label{sec:fault_tolerance}
The middleware is located between clients and cache nodes, and acts as a transparent proxy to aggregate connections for improving scalability, and to perform failure detection and data replication for improving fault-tolerance. In the next subsections, we provide technical background on the three middleware platforms analyzed in this study, with emphasis on their architecture and their tunable configuration.

\subsection{Twitter Twemproxy}

Twemproxy (aka Nutcracker) is a fast and light-weight proxy for the Memcached protocol. Twemproxy was developed by Twitter to reduce open connections towards cache nodes, by deploying a local proxy on every front-end node. Thanks to protocol pipelining and sharding, Twemproxy improves horizontal scalability of distributed caching. Twemproxy allows to create a pool of cache servers to distribute the data and to reduce the amount of connections. To send a request to any of those servers, the client contacts Twemproxy, which routes the request to a Memcached
server. The same Twemproxy instance can manage different pools, by setting different listening ports to each pool. The number of connections between the client and Twemproxy, and between Twemproxy and each server is set through a configuration file. It is interesting to note that the ``read my last write'' constraint does not necessarily hold true when Twemproxy is configured with more than one connection per server.

Twemproxy shards data automatically across multiple servers. To decide the destination server for a request, a \emph{request hashing} function is applied to the key of the request to select a shard; then, based on a chosen random distribution, the request is routed to a Memcached server in the shard. Twemproxy provides 12 different hash functions (i.e., \texttt{one\_at\_a\_time}, \texttt{MD5}, \texttt{CRC16}, \texttt{CRC32}, \texttt{FNV1\_64}, etc.) and 3 different distributions (i.e., ketama, modula, random). It allows the application to use only part of the key (\emph{hashtag}) to calculate the hash function. When the hashtag option is enabled, only part of the key is used as input for the hash function. This option allows the application to map different keys to the same server, as long as the part of the key within the tag is the same.

When a failure occurs, Twemproxy provides a mechanism to exclude the failed server. %, but since the replication is not managed data loss is unavoidable. 
It detects the failure after a number of failed requests (set in a configuration file), ejects the failed server from the pool, and redistributes the key among the remaining servers. Client requests are routed to the active servers; at regular intervals, some requests are sent to the failed server to check its status. When the failed server returns active, it is added back to the pool and the key is distributed again. Twemproxy increases observability using logs and stats. Stats can be at the granularity of server pool or individual servers, through a monitoring port. In the configuration file, it is possible to define the monitoring port and the aggregation interval. Companies that use Twemproxy in production include Pinterest, Snapchat, Flikr and Yahoo!.

\subsection{Facebook Mcrouter}

Mcrouter is a Memcached protocol router to handle traffic to, from, and between thousands of cache servers across dozens of clusters, distributed at geographical scale. It is a key component of the cache infrastructure at Facebook and Instagram, where Mcrouter handles nearly 5 billion requests per second at peak. It uses the standard ASCII Memcached protocol to provide a transparent layer between clients and servers, and adds advanced features that make it more than a simple proxy.

In the Mcrouter terminology, a Memcached server is a \emph{destination}, and a set of destinations is a \emph{pool}. Mcrouter adopts a client/server architecture: every Mcrouter instance communicates with several pools, without any communication between different pools. Several clients can connect to a single Mcrouter instance and share the outgoing connections, reducing the number of open connections. Looking at the Facebook infrastructure at a high level, one or more pools with Mcrouter instances and clients define a \emph{cluster}, and clusters together create a \emph{region}. Mcrouter is configured with a graph of small routing modules, called \emph{route handles}, which share a common interface (route a request, return a reply), and which can be combined. The configuration can be changed online to adapt routing to a transient situation, such as adding and warming up a new node. %Mcrouter checks for any changes in its configuration file, when it finds them, it reloads the file and adjusts the configuration. 
Since the configuration is checked and loaded by a background thread, there is no extra latency from the client point of view.

Mcrouter supports sharding in order to adapt the datastore tier to the growth of data. The data distribution among servers is based on a key hash. In this way, different keys are evenly distributed across different destinations, and requests for the same key are served by the same destination. It is possible to choose between different hash functions (e.g., \texttt{CH3}, weighted \texttt{CH3}, or \texttt{CRC32}). Keys in the same pool compete for the same amount of memory, and are evicted in the same way. 
Mcrouter also provides a feature, namely \emph{prefix routing}, that allows applications to control data distribution on different pools, by using different key prefixes. This feature is valuable since applications generate and store data of different complexity (e.g., caching the results of complex computations), and they can prevent cache misses on ``expensive'' data by using different prefixes than ``cheap'', so that they do not compete for the same memory space.
%At pool level, it is possible to distribute data by sending keys with different prefixes to different pools. 
%This is useful since Memcached makes no difference and doesn't prioritize any key. All keys compete for the same amount of memory and are evicted in the same way. 

Mcrouter supports data replication. Read and write requests are managed in a different way: writes are replicated to all hosts in the pool, while reads are routed to a single replica, chosen separately for each client, to achieve a higher read rate. %This might occur either because of a high read rate that cannot be managed by a single node or to increase the availability of the data. 
To increase fault tolerance, Mcrouter also provides destination health monitoring and automatic failover. When a destination is marked as unresponsive, the incoming requests will be routed to another destination. At the same time, health check requests will be sent in the background, and as soon as a health check is successful, Mcrouter will send the requests to the original destination again. Mcrouter distinguishes between \emph{soft errors} (e.g., data timeouts), which are tolerated to happen a few times in a row, and \emph{hard errors} (e.g., connection refused), which cause a host to be marked unresponsive immediately.
The health monitoring parameters are set up through command line options. It is possible to define the number of data timeouts to declare a \emph{soft TKO} (``technical knockout''), the maximum number of destinations allowed to be in soft TKO state at the same time, and the health check frequency. The health check requests (\emph{TKO probes}) are sent with an exponentially increasing interval, whose initial and maximum length (in ms) is configurable. The actual intervals have an additional random jitter of up to 50\% to avoid overloading a single failed host with TKO probes from different Mcrouters.

Mcrouter provides mechanisms to check the current configuration. Through an admin request, it is possible to verify what is the route of a specifc request. Given the operation and the key, the response from Mcrouter will be the server that owns the data. It is possible to get the route handle graph that a given request would traverse. These requests allow administrators to get insights on the actual state of the system, e.g., to show how the routing changes when a failure occurs.

\subsection{Netflix Dynomite}

Dynomite is a middleware solution implemented by Netflix. The main purpose of Dynomite is to transform a single server datastore into a peer-to-peer, clustered and linearly scalable system that preserves native client/server protocols of the datastores, such as the Redis and Memcached protocols. 

A Dynomite cluster consists of multiple \emph{data centers} (DC). A data center is a group of \emph{racks}, and a rack is a group of \emph{nodes}. Each rack consists of the entire dataset, which is partitioned across several nodes in that rack. A client can connect to any node on a Dynomite cluster when sending a request. If the node that receives the request owns the data, it gives the response; otherwise, the node forwards the request to the node that owns the data in the same rack. Dynomite is designed to be a sharding and replication layer. Sharding is achieved as in distributed hash tables \cite{chi2017hashing,tanenbaum2007distributed}: 
%Each node in a rack has a unique \emph{token} (an integer value), which helps to identify the data it owns. 
Request keys are hashed to obtain a \emph{token} (an integer value); the range of all possible token values is partitioned among the nodes, by assigning to each node a unique token value; the keys that fall in the sub-range ending with the node's unique token are assigned to that node. 

To achieve replication, datasets are replicated among all the racks. When a node receives a \texttt{set} request, it acts as a coordinator, by writing data in the node of its rack which owns the token, and by forwarding the request to the corresponding nodes in other racks and data centers. 
Every node knows the distribution of the token in the nodes of all data centers through to the configuration file. Using a similar technique for \texttt{get} requests, it is possible to increase data availability and tolerate node failures. The configuration file defines the topology, the number of connections and the token range partitioning among the nodes. Moreover, it is possible to configure one of three confidence levels writes and reads: 

\begin{itemize}

\item \textbf{DC\_ONE}: requests are propagated synchronously only in the local rack, and asynchronously replicated to other racks and regions;

\item \textbf{DC\_QUORUM}: requests are sent synchronously to a quorum of servers in the local data center, and asynchronously to the rest. This configuration writes to a set of nodes that forms a quorum. If responses are different, the first response that the coordinator received is returned;

\item \textbf{DC\_SAFE\_QUORUM}: similar to DC\_QUORUM, but data checksums have to match.

\end{itemize}

Dynomite makes the caching system tolerant to failures of one or few servers (i.e., less than the size of the quorum) through replication, but it does not provide any failover detection or online reconfiguration. This makes the system less dynamic. For example, to add a new node you have to recompute all tokens, update the configuration file, and restart all processes. Failover detection, token reconfiguration, and other features are provided from other components in the Dyno ecosystem, like Dynomite-manager or Dyno client. Dynomite provides REST APIs for checking the state of some parameters (e.g., the confidence level), but this does not include routing (i.e., where a key is stored or read from).

\subsection{Qualitative comparison of fault-tolerance mechanisms}

Both Mcrouter and Dynomite adopt replication as strategy to achieve transparent fault tolerance. Of course, this strategy has a major impact on performance, since a single request turns in multiple operations on several machines. Given the same settings (number of clients, connections, threads, etc.), the throughput of the caching system can significantly degrade when enabling replication: for example, if throughput without replication reaches $20,000$ reqs/sec, it can settle to $5,500$ reqs/sec when using a replication factor of three. 
%Without replication, throughput is higher but a failure is not completely masked; with replication, lower performance is compensated by a complete transparency of the fault.
%This makes even clearer how the replication leads to a decrease in performance since a single request turns into reality in more operations on the machines.
The same is not necessarily true for latency, which is of the same order of magnitude in all cases. Being able to use replication without affecting latency is the reason why major service providers are using this strategy despite lower throughput, since latency is the key performance indicator that mostly influences the Quality of Experience.

Mcrouter has a more complete failure management, as compared to the previous ones. It consists on monitoring the health state of destinations and detecting failures. By default, it converts all \texttt{get} errors into cache misses, so that the application can still work even if there is no replication and a high amount of errors occur. %It is interesting to note that a high amount of errors results in a failure of the whole system, while the system continues to work if there is a high amount of misses. So, this mechanism is useful when is not provided a fault-tolerant configuration.
Similar considerations on misses and errors also apply for Twemproxy. When  keys are redistributed among nodes after a failure, Twemproxy turns the timeouts into cache misses. 

Both Twemproxy and Mcrouter rely on knowledge about the infrastructure and its failures to be configured with appropriate values, e.g., for the frequency of monitoring, the timeouts, and the replication policies. The main factors driving the configuration are: the fault duration, the throughput, and the amount of data on each server. For example, if we have a throughput of 20k \texttt{set} reqs/sec, and 100k keys on each node, then all keys can be redistributed in 5 seconds and be again available. Thus, it is worth adopting redistribution when the throughput is high, the amount of data on each node is small, and the outage of faulty nodes is expected to b long.

Dynomite is fault-tolerant with some limitations. In particular, if the coordinator node fails (i.e., the node that first receives a request), there is no retry mechanism that sends the request to another node. Thus, coordinator failures are not tolerated. This is a deliberate design choice, since Dynomite is designed to be on the same host running a Memcached server, thus when the host fails, both Memcached and Dynomite fail. Furthermore, if the rack includes a failed node, the rack still continues to count towards the quorum, while the rack is excluded from the quorum when all its nodes are failed.

\section{Experimental methodology and setup}
\label{sec:experiment}
% !TEX root = ../main.tex
% !TEX encoding = UTF-8 Unicode
Our experimental campaign aims to identify potential dependability issues that emerge when caching systems are deployed at a large scale. We deployed the three middleware platforms on a testbed with dozens of caching nodes, and we adopted \emph{fault injection} to accelerate the occurrence of faults for testing purposes. Fault injection artificially injecting faults that emulate common failure scenarios in a large-scale distributed system.

In this study, the target system is composed by two tiers: (i) the \emph{Middleware} tier, which includes nodes running the middleware platform under analysis, by accepting clients' requests and forwarding them to Memcached servers; and (ii) the \emph{Cache} tier, which contains Memcached servers where the actual data are stored and fetched. We focused our analysis on Memcached since it is widely used by many internet companies  such as Facebook \cite{xu2013characterizing}, Google \cite{lee2014cache}, and many others \cite{pinterest2013building,rajashekharcaching,papapanagiotou2018ndbench} and it is well supported by the middleware software under analysis.

\begin{figure}[ht]
    \centering
    \includegraphics[width=\columnwidth]{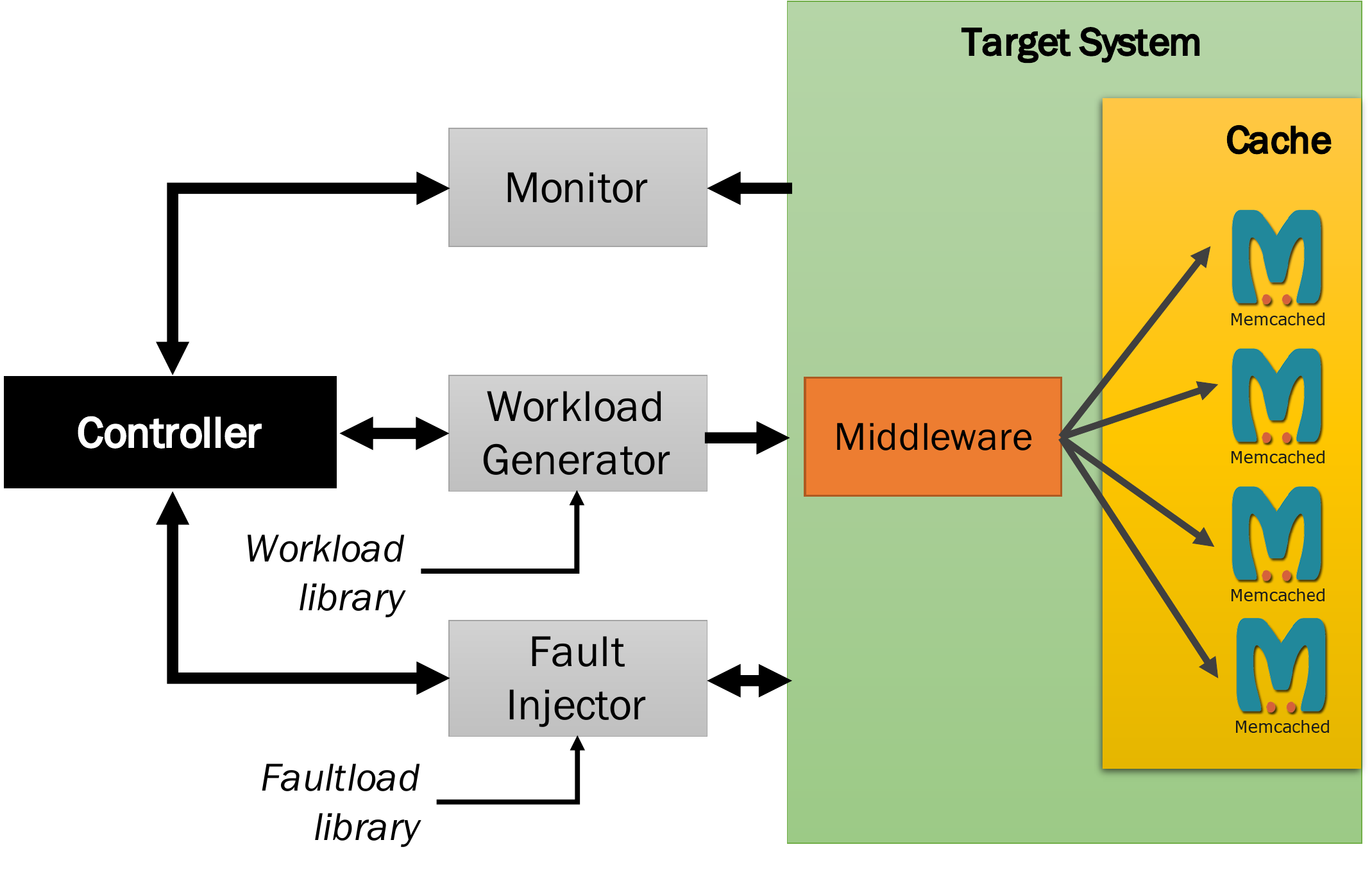}
    \caption{Fault injection testbed} 
    %\hl{PER MIGLIORARE LA QUALITÀ DELLA FIGURA, SI PUO AUMENTARE IL FONT, LA DIMENSIONE DELLE FRECCE, ENFATIZZARE I BORDI DEI BOX}
    \label{fig:architecture}
\end{figure}

In addition to these components, our testbed includes a fault injection environment with the following components:

\begin{itemize}

\item \textbf{Fault injector}: It injects faults into the target system during the experiment. It can be custom-built hardware or software. In our environment, we use a software fault injector implemented through a bash shell script which kills and overloads the victim nodes.

\item \textbf{Workload generator}: It emulates user traffic to exercise the target system, e.g., by running actual applications, benchmarks, or synthetic workloads. In our environment, it is implemented through a Python application, as discussed in the following.  The experiment data concerning database requests (e.g. throughput, latency, misses, etc.) are also collected by this component.

\item \textbf{Monitor}: It performs data collection, and keeps track of the progress of experiments. In our environment, we use a bash shell script. As we use Memcached server for the caching tier, we monitor the caching servers, including metrics such as total cache size, global hit percent, number of connections, \texttt{get}/\texttt{set}/\texttt{delete} requests per second, and network bandwidth utilization. Moreover, we collect global statistics for the clusters as a whole and for individual servers, including slabs, occupation, memory wasted, and items stored in a server \cite{fitzpatrick2004distributed}.  The monitor also collects data about resource usage (e.g. CPU, memory, net, etc.). This data is retrieved by daemons installed in each virtual machine of the target system (including both the database instances and the middleware VMs) which periodically read system performance metrics from the proc filesystem.

\item \textbf{Controller}: It orchestrates the experiments. It can be a program that runs on the target system or on a separate computer. In our testbed, it is implemented through a bash shell script and it is located in a dedicated host. Its tasks include: resetting the machines before starting the experiment, warming up the servers, starting the workload generators, and triggering fault injection.

\end{itemize}

%Faults can be hardware/software residing in Memcached systems and Congestion problems due to the system architecture. In this context 
We have considered three different scenarios that can affect large-scale distributed systems:

\begin{enumerate}

    \item \textbf{Crash failure of Memcached servers}. This is the classical fault model adopted in fault injection for distributed systems, where a node is shut down and unresponsive because of hardware or software faults. Examples include power outages, broken CPU or RAM components (e.g., due to over-heating), logically- or physically-severed network connections, security attacks (e.g., a DDoS), or software crashes due bugs in the OS or applications. The fault injector emulate crashes by sending UNIX signals to Memcached servers in order to stop and to restart them.
    
    \item \textbf{Unbalanced overloads of a subset of Memcached servers}. 
    This represents a so-called gray failure. Overload conditions (i.e., insufficient resources to serve the incoming user traffic) are a critical aspect for the design of large-scale distributed applications \cite{Bauer2012RAC}. \emph{Unbalanced} overloads are especially problematic: they occur when a small subset of overloaded nodes become a bottleneck for the whole distributed system. One typical case is represented by \emph{hot-spots}, e.g., popular multimedia resources and application features \cite{hong2013understanding, huang2014characterizing, oonhawat2017hotspot}. Unbalanced overloads can also be caused by \emph{over-commitment} (e.g., the same host is shared by multiple tentants, which face a high load at the same time) and design faults in system configuration and software \cite{cotroneo2019overload}. These faults are often overlooked in the design of distributed systems, as they cause subtle effects that are difficult to detect. Thus, our experiments evaluate how the middleware platforms contribute to mitigate (or, exacerbate) unbalanced overloads in the caching nodes. 
    This fault is emulated by introducing CPU-bound processes on a subset of Memcached nodes.
    
    \item \textbf{Link bandwidth bottlenecks}. Network faults can affect links, switches, and NICs in a data center. Similarly to unbalanced overloads, the worst case is represented by silent, non-fail-stop faults in small parts of the network. Typically, these faults are caused by wear-out of NICs and links, which lead to packet loss and performance degradation \cite{gunawi2018fail}. These faults are emulated by re-configuring the OS to restrict the bandwidth of network interfaces (e.g., from 1000 Mbps to 100 Mbps).
    
\end{enumerate}

\subsection{Workload generation}

To generate a workload for the caching nodes and for the middleware platform, we need i)to create connections to send data requests, and ii) to collect performance metrics from the clients' perspective. We initially tried several publicly-available tools, but many proved not to be suitable for carrying out our experiments. 
The first tool was the \emph{CloudSuite Data Caching Benchmark} \cite{cloudsuiterepo}, which is part of CloudSuite, a benchmark suite for cloud services. This tool does not support execution in the presence of failures (which we will deliberately force with fault injection), and stops making requests when a server fails. 
Then, we tried \emph{Memaslap} \cite{memaslap}, a load generation and benchmark tool for Memcached servers included in \emph{libMemcached}. However, the format of requests generated by this tool are incompatible with Mcrouter. 
For this reason, we developed our own tool, namely Mcrouter \emph{Mcbench}. It is written in Python, uses the \emph{aiomcache} library to create the connections and makes requests according to the Memcached protocol. Mcbench generates only \texttt{set} (20\%) and \texttt{get} (80\%) requests with random keys and values. %To generate a request, the tool selects a random key from a distribution configured by the user. For \texttt{set} requests, it also generates a value with a random length, based on a distribution associated to the key.
The tool can be configured with respect to the distribution of keys and values lengths; 
the duration of the experiment; 
the desired throughput; 
the number of threads, clients and connections per server; 
the list of servers where to forward requests.

The tool collects performance statistics about requests that are successful and failed, latency, and misses, aggregating them every second. %It writes these data in a csv file used in the analysis phase. 
To warm up servers, we configure the tool to initially perform only \texttt{set} requests. The random distribution of the keys is based on a real-world Twitter dataset from the CloudSuite Data Caching Benchmark \cite{twitterdataset}. To create random distributions for a more intensive workload for large distributed systems, we scaled-up the Twitter dataset by a factor of ten, while preserving both the popularity and the distribution of object size, with alphanumeric variable-length keys. 
The original dataset consumes 300MB (267,433 rows) of server memory, while the scaled dataset requires about 3GB (2,674,339 rows).

\subsection{Testbed configuration}

The experimental testbed was built on top of OpenStack, an open-source Infrastructure-as-a-Service (IaaS) cloud computing platform. 
The experimental testbed consists of eight host machines SUPERMICRO (high density), equipped with two 8-Core 3.3Ghz Intel XEON CPUs (32 logic cores in total), 128GB RAM, two 500GB SATA HDD, four 1-Gbps Intel Ethernet NICs, and a NetApp Network Storage Array with 32TB of storage space and 4GB of SSD cache. The hosts are connected to three 1-Gbps Ethernet network switches, respectively for management, storage and VM network traffic. The testbed is managed with OpenStack Mitaka and the VMware ESXi 6.0 hypervisor. 
%First and foremost it is necessary to instantiate  virtual machines on Openstack in order to reach the scale of the experiments. In this phase of experimentation, the testbed is composed of the virtual machines described in \hl{Table}.

\begin{table}
\centering
\protect\caption{Configuration of the experimental testbed}
\label{tab:configuration}

\begin{tabular}{|p{1.5cm}|p{0.6cm}|p{5cm}|}
\hline 
\textbf{Node Type} & \textbf{\# of nodes} & \textbf{VM configuration}\tabularnewline
\hline 
\hline 
\textbf{Controller Node} & 1 & 2 vCPU, 16GB RAM, 120 GB HDD\tabularnewline
\hline 
\textbf{Middleware Node}  & 7 & 4 vCPU, 4 GB RAM, 20 GB HDD\tabularnewline
\hline 
\textbf{Memcached Node} & 30 & 1 vCPU, 8 GB RAM, 20 GB HDD\tabularnewline
\hline 
\textbf{Workload generator}  & 7 & 4 vCPU, 8 GB RAM, 40 GB HDD\tabularnewline
\hline 
\textbf{\emph{Total}} & \emph{45} & \emph{88 vCPU, 340 GB RAM, 1.14 TB HDD}\tabularnewline
\hline 
\end{tabular}

\end{table}

The system architecture detailed in Table \ref{tab:configuration} consists of: 
\begin{itemize}
\item one controller machine, which orchestrates the test and sends commands to the other machines; 
\item seven workload generators, which send requests to the middleware platform;
\item seven middleware machines, which forward the requests to the Memcached servers;
\item thirty Memcached servers.
\end{itemize}

\begin{figure}[ht]
    \centering
    \includegraphics[width=\columnwidth]{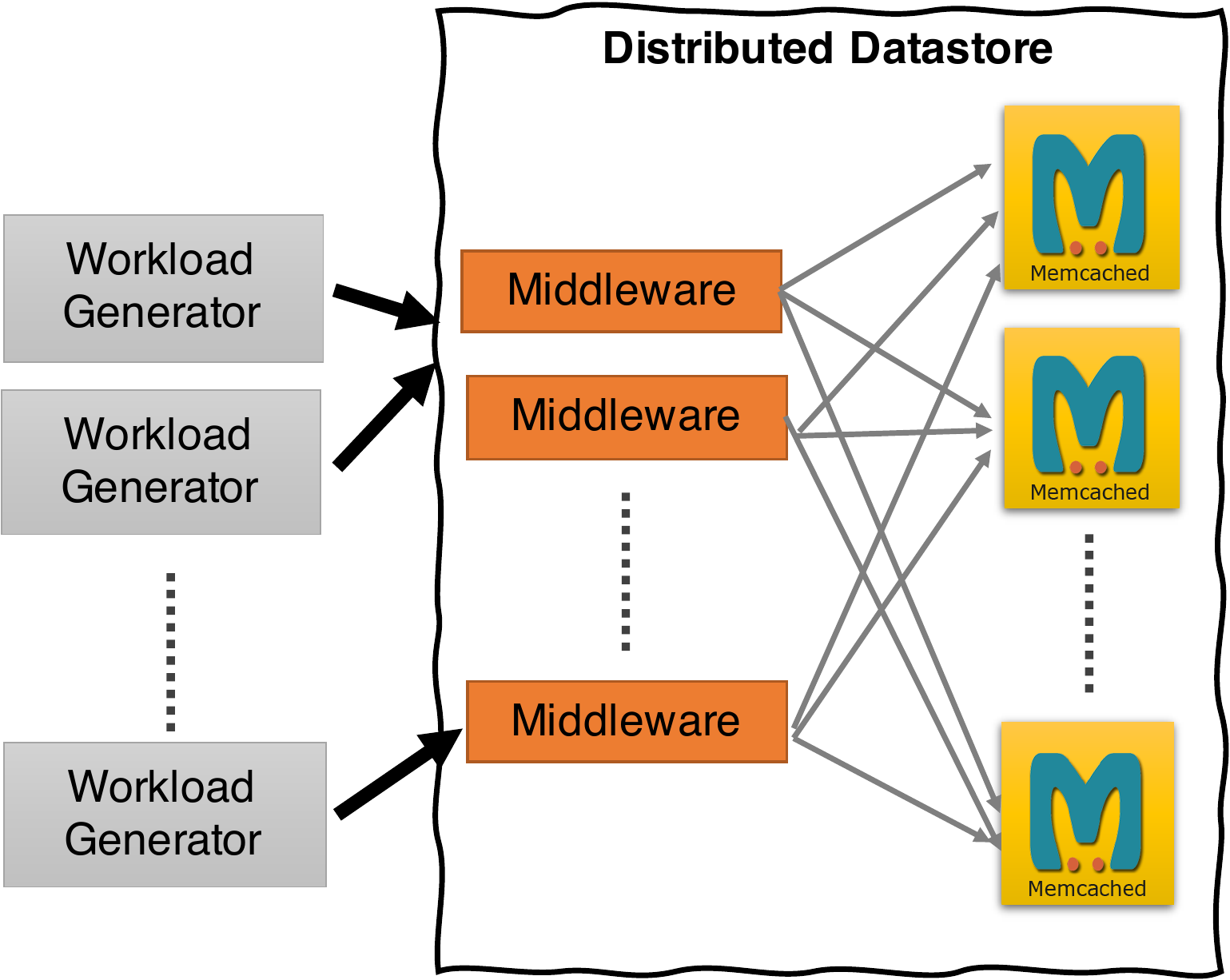}
    \caption{Twemproxy and Mcrouter topology}
    \label{fig:topology1}
\end{figure}

\begin{figure}[ht]
    \centering
    \includegraphics[width=\columnwidth]{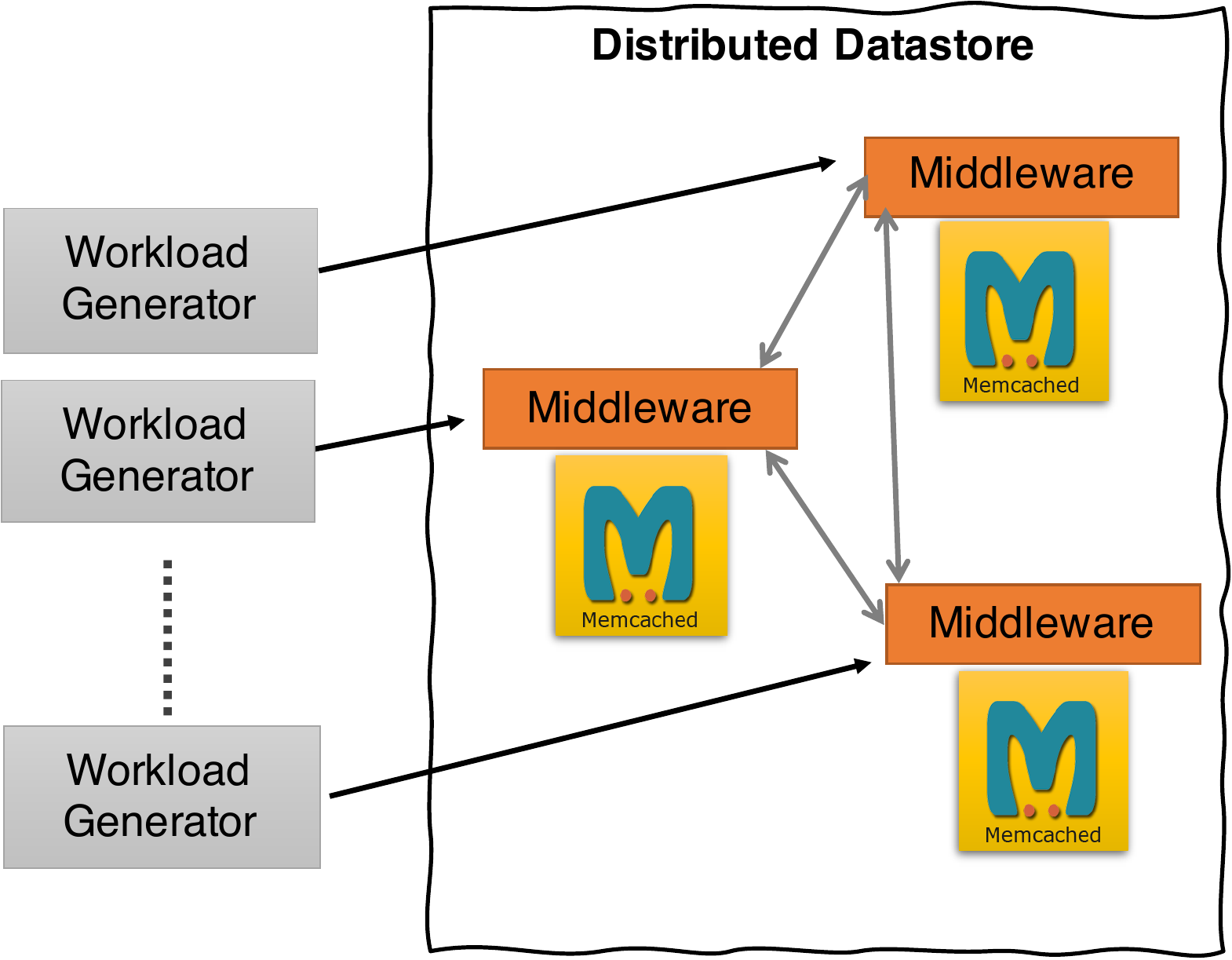}
    \caption{Dynomite topology}
    \label{fig:topology2}
\end{figure}

The connections between the middleware and the Memcached servers vary depending on the specific middleware. In the case of Twemproxy and Mcrouter, each instance of the middleware connects to all servers, as in \figurename{}~\ref{fig:topology1}. In Dynomite, each instance connects only to one node, since there as many Dynomite instances of Dynomite as the number of Memcached servers, as shown in \figurename{}~\ref{fig:topology2}. Therefore, in Dynomite there are fewer connections between the middleware and the Memcached servers, and the instances of the middleware communicate with each other to route the requests.% by establishing connections, which in the case of Mcrouter and Twemproxy does not happen.


To have a balanced load among the physical hosts, we equally divided the VMs in order to have one workload generator, one middleware and at most five Memcached servers on each physical host. 
For Twemproxy, the dataset is divided among all thirty nodes; for Mcrouter and Dynomite, which support replication schemes, we have three pools/racks, each composed of ten nodes among which the dataset is divided, thus with a replication factor of three. Thus, a $set$ operation is performed on three different nodes and a get operation in one node out of three. In case of a miss or of an error the operation is retried on the remaining nodes and it is returned the first valid response if any. If the middleware supports the quorum (i.e., in the case of Dynomite) the middleware should reply with an error if the reply does not reach the quorum of two valid responses.
We carried out a capacity test to define the maximum throughput that can be reached by the system. The maximum throughput is $18,500$ reqs/sec per workload generator with Twemproxy, and $7,200$ reqs/sec per workload generator with Mcrouter and Dynomite. We then calibrated the workload generator to have a throughput per workload generator of $16,500$ reqs/sec for Twemproxy, and $6,500$ reqs/sec for Mcrouter and Dynomite. The total number of requests served in the experiments are respectively $115,500$ reqs/sec and $45,500$ reqs/sec.

\section{Experimental results}
\label{sec:discussion}
In the following, we report on the results of fault injection experiments, with three subsections for each category of faults. Each experiment was repeated five times. We obtained very low variations across different runs on the aggregated data across the whole cluster (throughput, latency, misses and errors). We carefully analysed data to see if all runs demonstrate the same failure behavior and choose a representative one out of the five to present the results.

\subsection{Crash of Memcached servers}
%CRASH

%The crash of a server is one of the most intuitive faults that can be considered for a system like the one under analysis, as well as a certain event in the long term. Indeed, a system that has a 100\% availability is a utopia, in practices, a high availability system is 5 nines available, i.e. 99.999\%, a safety critical systems reach 6 nines. Therefore servers definitely have downtime periods, whose causes are very heterogeneous. Downtime events typically arise from some physical event, such as a hardware or software failure or environmental anomaly. Examples of unscheduled downtime events include power outages, failed CPU or RAM components (or possibly other failed hardware components), an over-temperature related shutdown, logically or physically severed network connections, security breaches, or various application or operating system failures. 

In these experiments, Memcached is stopped on a subset of the caching server nodes. An experiment has a total duration of 600 seconds. We have an initial phase of 200 seconds (\emph{warm-up}), in which the system and the workload execute without faults; then, we have a second phase of 200 seconds (\emph{fault-injection}), in which the fault is actually injected in the system; and a final phase of 200 seconds (\emph{recovery}), in which the fault is removed from the system. Several experiments were performed by increasing the number of failed servers, to evaluate how this affects the behavior of the system and its fault tolerance mechanisms.

\begin{figure}[ht]
    \centering
    \includegraphics[width=\columnwidth]{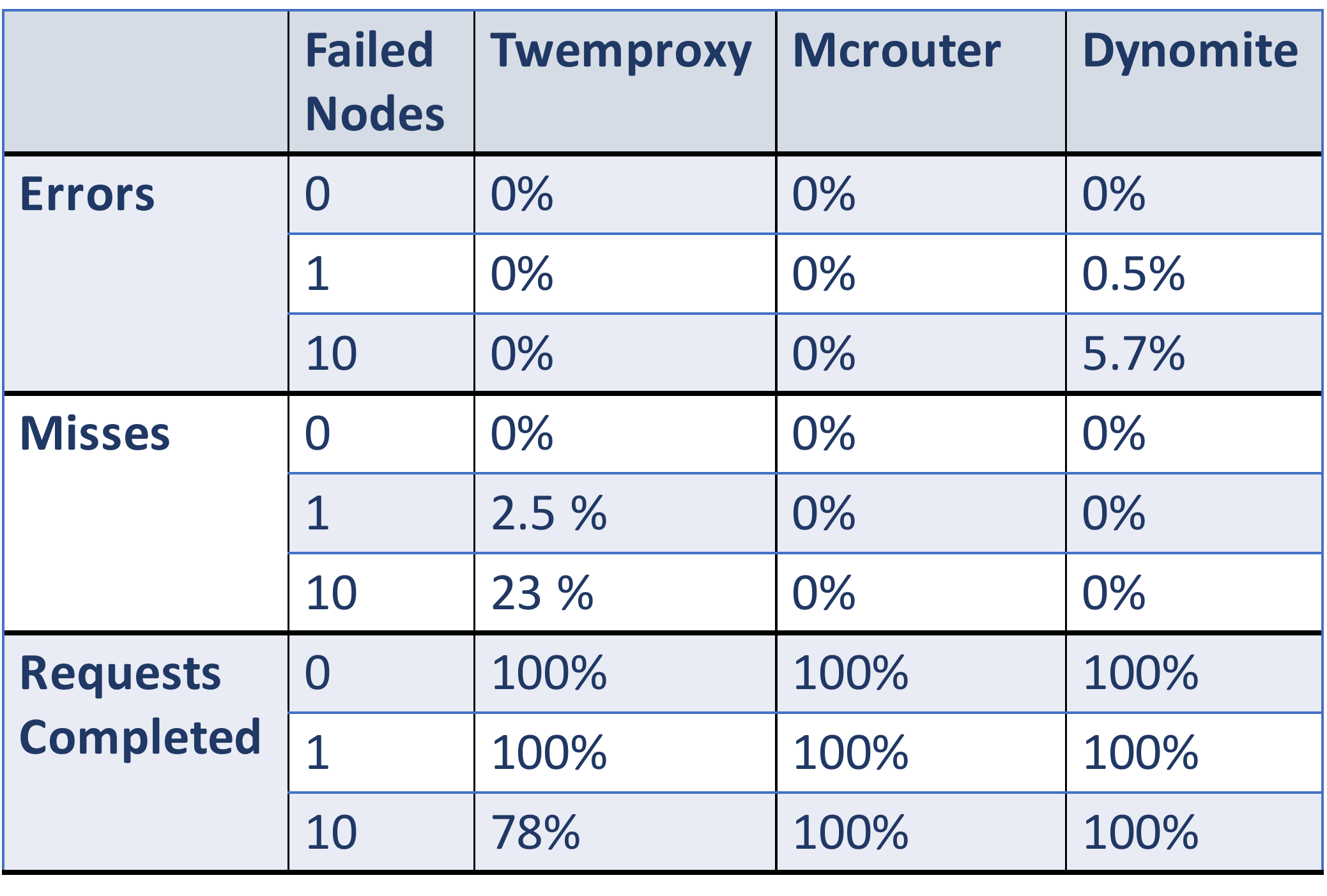}
    \caption{Performance under Memcached crash failures}
    \label{fig:node_failure_results}
\end{figure}

When Memcached nodes crash, the three middleware platforms react in different ways. Client requests can experience three possible outcomes, as summarized in \figurename{}~\ref{fig:node_failure_results}: \emph{done} (i.e., the key-value pair is correctly set or retrieved), \emph{misses} (i.e., the key-value pair cannot be retrieved because of the fault, despite it is actually stored by the caching tier); \emph{errors} (i.e., the middleware returns an error signal to the clients). 

\begin{figure}[ht]
    \centering
    \includegraphics[width=\columnwidth]{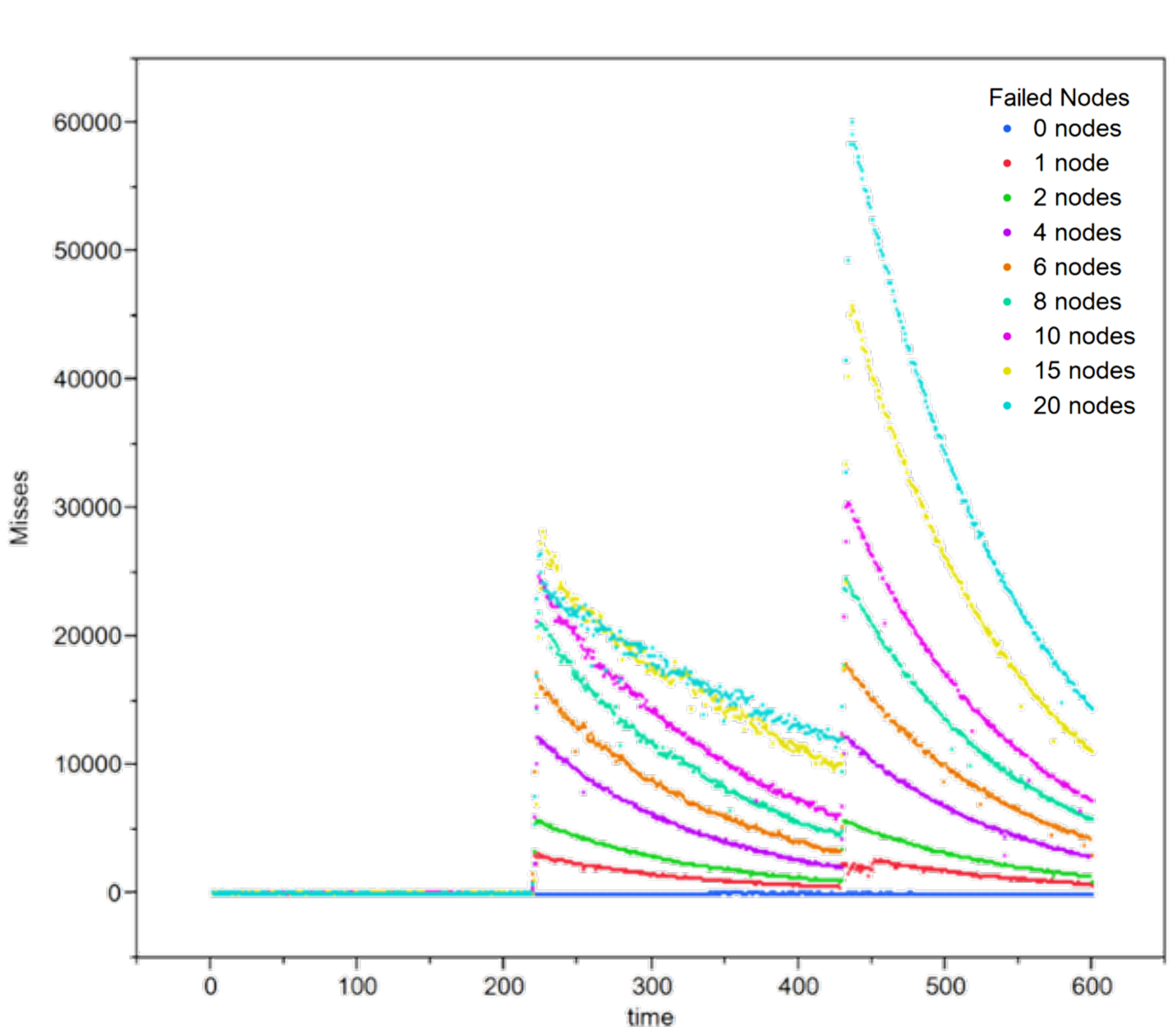}
    \caption{Cache miss rate during experiments with crash of Memcached servers}
    \label{fig:misses}
\end{figure}

By using Twemproxy, we have cache misses both during the failure and the recovery phase (two values are reported in the cells of the table). Moreover, when the number of failed server is higher than 4, we observe a significant throughput decrease and latency increase, due to the timeouts required to activate the node ejection mechanism. We found that cache misses during the recovery phase are even more than during the fault-injection phase due to a redistribution of the keys in the nodes that are still alive.  \figurename~\ref{fig:misses} shows in detail this behaviour: after 200s a subset of memcached nodes are stopped and restarted after 200s. During the fault, the part of the keys are redistributed in the remaining nodes. Those keys are potential future cache misses during the recovery phase due to another change in the topology. This behavior happens because Twemproxy does not provide mechanisms for data replication that could mask data losses during node failures, and does not support online topology changes without redistributing keys among the nodes.

%TWEMPROXY
%- Node crash
%---> latency increases and throughput decreases wrt \# nodes failed (\#failure > 4)
%---> misses when fault is ON are less than misses when fault is again OFF (note: seems that keys are %redistributed somehow in case of failure, but there are no features for that in Twemproxy)

%MCROUTER

%- Node crash
%---> No misses/timeouts, no significant latency variation if only 1-out-of-3 servers with the same data is out
%---> If two nodes with the same data are out, there are some timeouts (during fault, the node up can be overloaded) and misses (after fault, the node that was up is the only one with the data, and can be overloaded, and requests are routed to the recovered nodes that do not have yet the data)
%---> latency is stable since unresponsive nodes are detected and removed;

Mcrouter and Dynomite both support failure detection and data replication. Under the same conditions, Mcrouter is able to mask Memcached failures without significant latency and throughput variation when only 1-out-of-3 servers with the same data are crashed. If two nodes with the same data are crashed, there are timeouts errors, since the only alive server becomes overloaded. The unresponsive nodes are promptly detected and removed from the topology and thus we did not found effects on the latency, which remained stable up to 10 out of 30 node failures. Dynomite is also capable to mask node crashes. However, due to the specific peer-to-peer architecture of Dynomite, when the failed node is in the same rack of the coordinator, and the received responses from other racks are different (such as a cache miss and a cache value), the coordinator always replies with the local rack response \cite{dynomiteConsistency} which is an error . Moreover, in case of two-out-of-three crashed nodes with the same data, all requests to these data experience a failure due to the impossibility to reach the quorum. Thus, differently from Mcrouter, the system does not suffer from an overload of the only available replica, but the clients can experience data unavailability (i.e., service errors).

%DYNOMITE

%- Node crash
%---> If data are replicated, and the failed node is in the same rack of the coordinator, the "quorum" mechanism returns the first response (which may be an error)
%---> When the failed node is recovered, there are misses
%---> If two-out-of-three nodes with same data are failed, almost all requests to that data experience a failure

\subsection{Unbalanced overloads of Memcached servers}

%An unbalanced load is one of the most critical problems in distributed systems. More than an actual fault, it is an unusual condition of the system's functioning, so in many cases, it is more difficult to identify and manage. 
In this scenario, a subset of the caching servers experiences an overload, which does not simply cause a fail-stop behavior (such as crashes, which are explicitly notified by the OS), but degrades the performance of the servers without explicit failure notifications. In all of the middleware platforms, the data and the requests are distributed among nodes using a stateless and fixed scheme, by computing a deterministic hash function on the keys (\emph{consistent hashing}). This approach selects a node (or a subset of nodes) only on the basis of the key, without considering the workload of the nodes across the datacenter. This can result in an uneven distribution of the workload, such as when most of the users' requests are on a specific resource (\emph{hot-spot}), or when there is physical resource contention on specific nodes. %We also have to consider that a subset of nodes can be overloaded because nodes have different capacity, thus their capacity/load ratio can vary over time, for example, due to physical resource contention, or due to faults.

\begin{figure}[ht]
    \centering
    \includegraphics[width=\columnwidth]{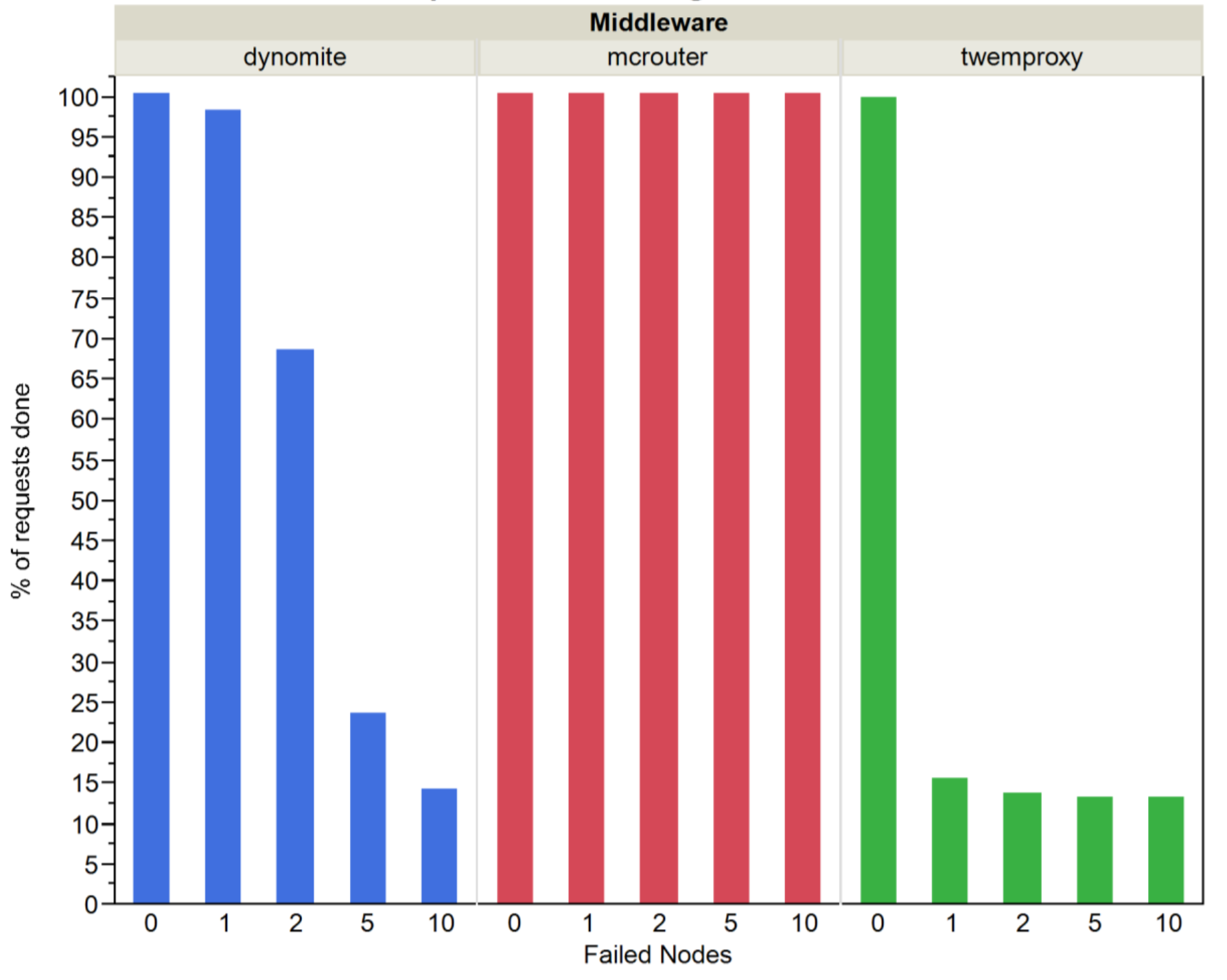}
    \caption{Performance under unbalanced overloads of Memcached nodes.}
    \label{fig:unbalanced_load_results}
\end{figure}

\figurename~\ref{fig:unbalanced_load_results} shows the percentage of requests done for the three middleware platforms, when injecting overloads on an increasing number of caching servers. 
In Twemproxy, the presence of unbalanced overloads significantly affects the performance, since these nodes are slowed-down, but are not enough to trigger the ejection mechanism and keys reallocation. It is interesting to note the absence of correlation between the number of failed servers and the number of timeouts errors. This means that having just one or several overloaded nodes has the same impact on performance. We looked at resource utilization metrics of the non-overloaded nodes, and found that these nodes are under-performing much below their normal load, despite they are not directly targeted by fault injection. 
This behavior occurred because most of the requests in the Twemproxy queue were waiting on the network sockets opened towards the overloaded nodes, and could not be quickly re-sent to another node. Therefore, the effects of one faulty node propagate through the middleware to the whole caching tier.

\begin{figure*}[ht]
    \centering
    \includegraphics[width=0.9\textwidth]{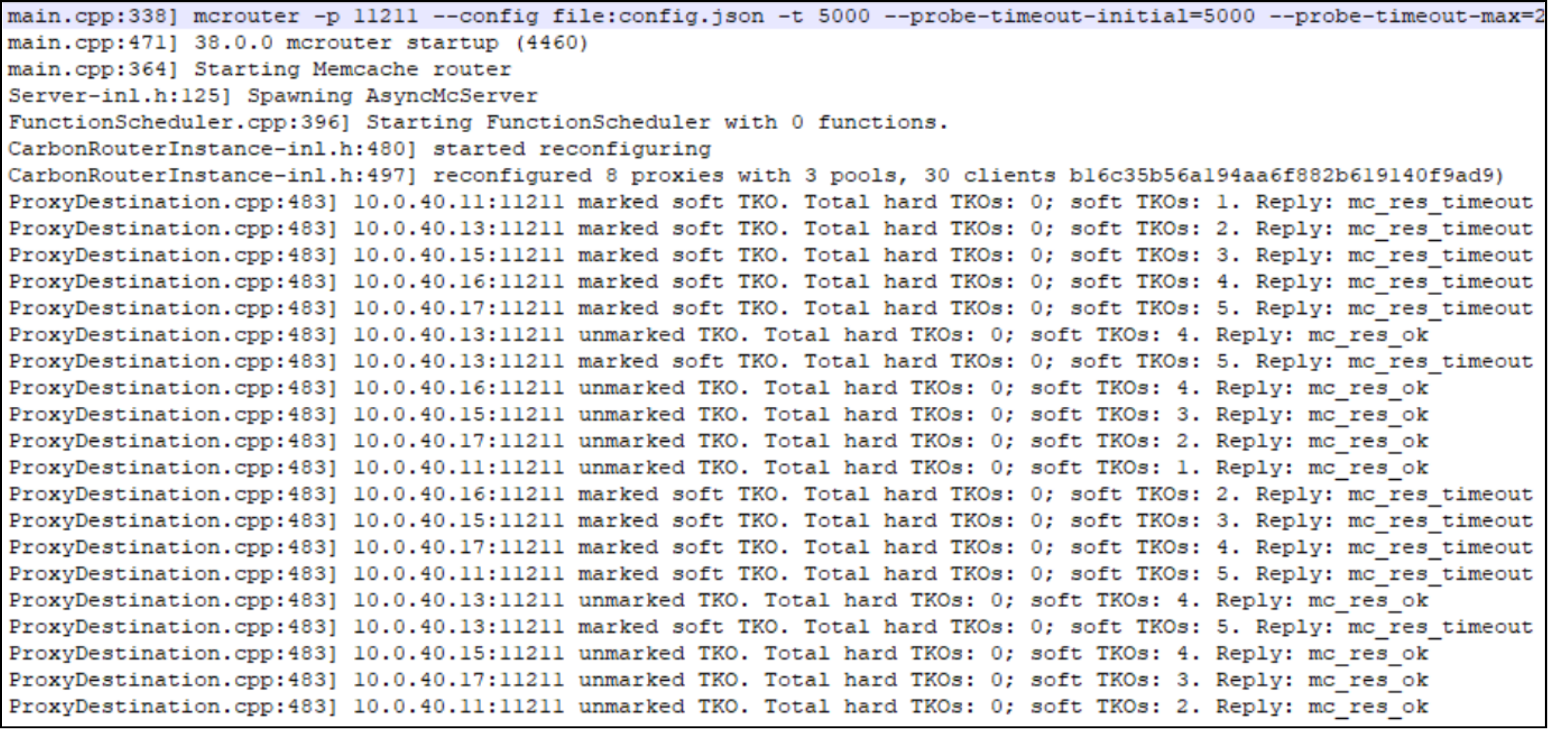}
    \caption{Mcrouter logs during unbalancing overloads.}
    \label{fig:logs}
\end{figure*}

Among the three middleware platforms, Mcrouter is the one most robust against unbalanced overloads, as there is no degradation of the rate of requests done (\figurename{}~\ref{fig:unbalanced_load_results}). The fault does not impact on the throughput and latency of the caching tier. Resource utilization remains stable in non-overloaded nodes, and only saturates on the overloaded ones. This behavior was due to the failover mechanism, which is able to detect the overloaded nodes, mark them as \emph{soft TKO} as shown in the logs in \figurename{}~\ref{fig:logs}, and remove them from the list of routes. Thus, Mcrouter addresses both overloads and crashes of servers in a similar way. Thanks to the exclusion of failed servers, requests are directly sent to another node without any waiting time, thus achieving a stable latency.  Regardless of the pool of the overloaded nodes (in the same pool, or in different ones), performance is not affected. 
Instead, when we inject an overload in two nodes with the same data, performance is inevitably degrades. In our experiments, throughput reduces by 85\% and  latency grows by two orders of magnitude. The overloaded nodes alternate between states (active vs. \emph{soft TKO}), with one node active and the other unavailable, and vice versa. %Therefore, for most of the time, there is a delayed node and a normal node that works at the same time. 
This behavior increases latency, since \texttt{set} requests cause the middleware to wait for responses from an overloaded node to achieve a majority. In the case of \texttt{get} requests, part of them are sent to an overloaded node.

\begin{figure*}[ht]
    \centering
    \includegraphics[width=0.9\textwidth]{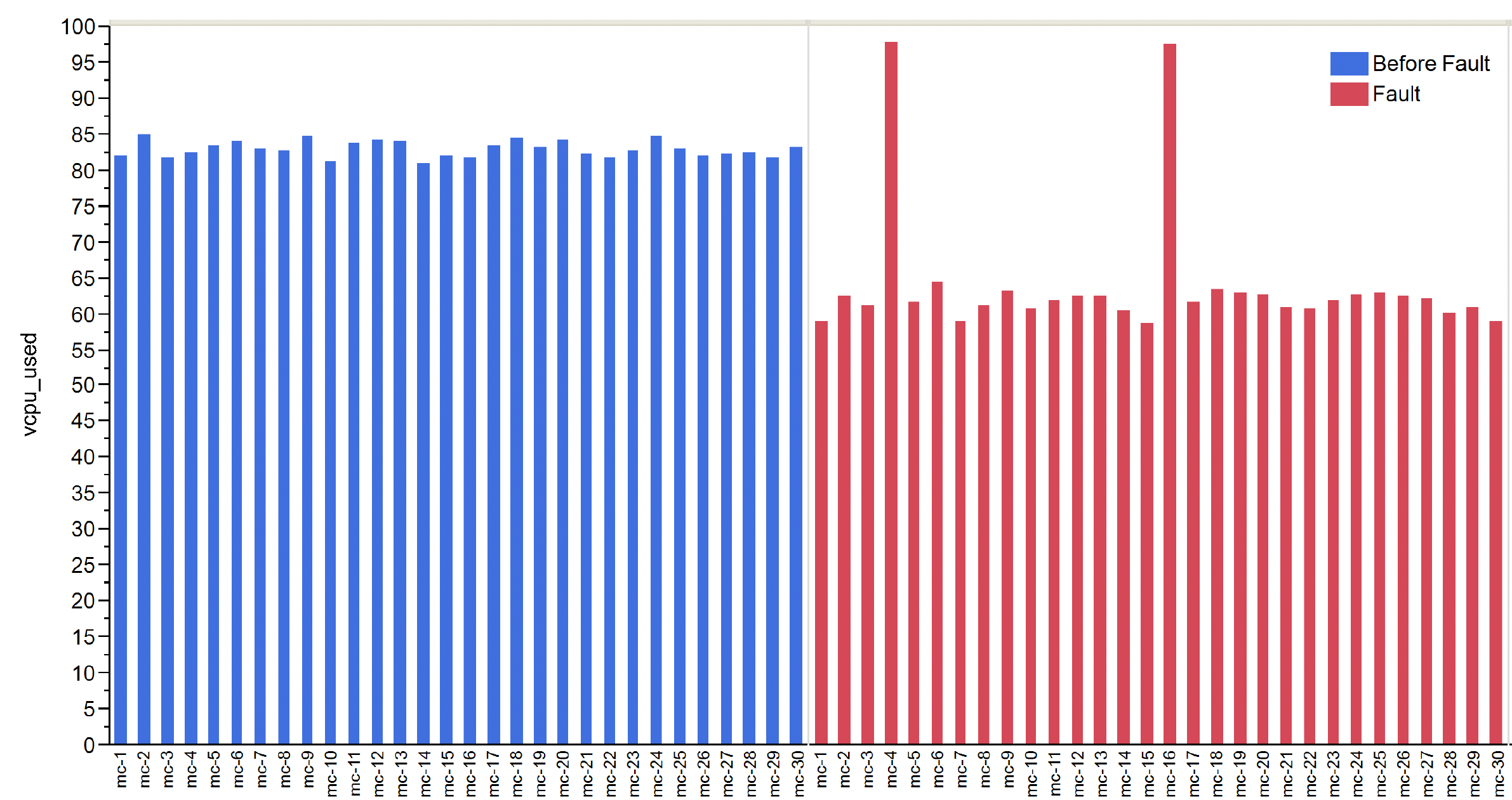}
    \caption{Average vCPU usage before (blue) and after (red) unbalanced overload injection.}
    \label{fig:dynomite_unbalanced_cpu}
\end{figure*}

Dynomite is not able to properly manage unbalanced overloads. The average throughput decreases proportionally with the increase of overloaded servers (\figurename{}~\ref{fig:unbalanced_load_results}). In this middleware platform, the throughput degrades due to the high latency of requests processed by the overloaded servers. \figurename~\ref{fig:dynomite_unbalanced_cpu} shows the effects of the overload of two out of thirty servers on the cpu usage of the remaining nodes. In particular we observe a significant cpu usage reduction across the whole datastore cluster. %In particular, requests concerning unbalanced nodes have higher latency that entails the reduction of throughput, whereas the others requests are faster and entail the increase of throughput. 
Requests for a given key can experience a high or low latency, depending on whether an overloaded node is chosen among the nodes of the quorum.  Differently from the Twemproxy case, Dynomite does not propagate the effect of overloaded nodes to the remaining servers.

\subsection{Network link bottlenecks}

In these experiments, we inject faults in the caching tier by reducing the network bandwidth of a group of servers. In this case, the fault affects a group of Memcached servers that are hosted on the same host, to reproduce a fault (e.g., network congestion) happening at the infrastructure level. Restricting the bandwidth leads to a bottleneck in the network. 

\begin{figure}[ht]
    \centering
    \includegraphics[width=\columnwidth]{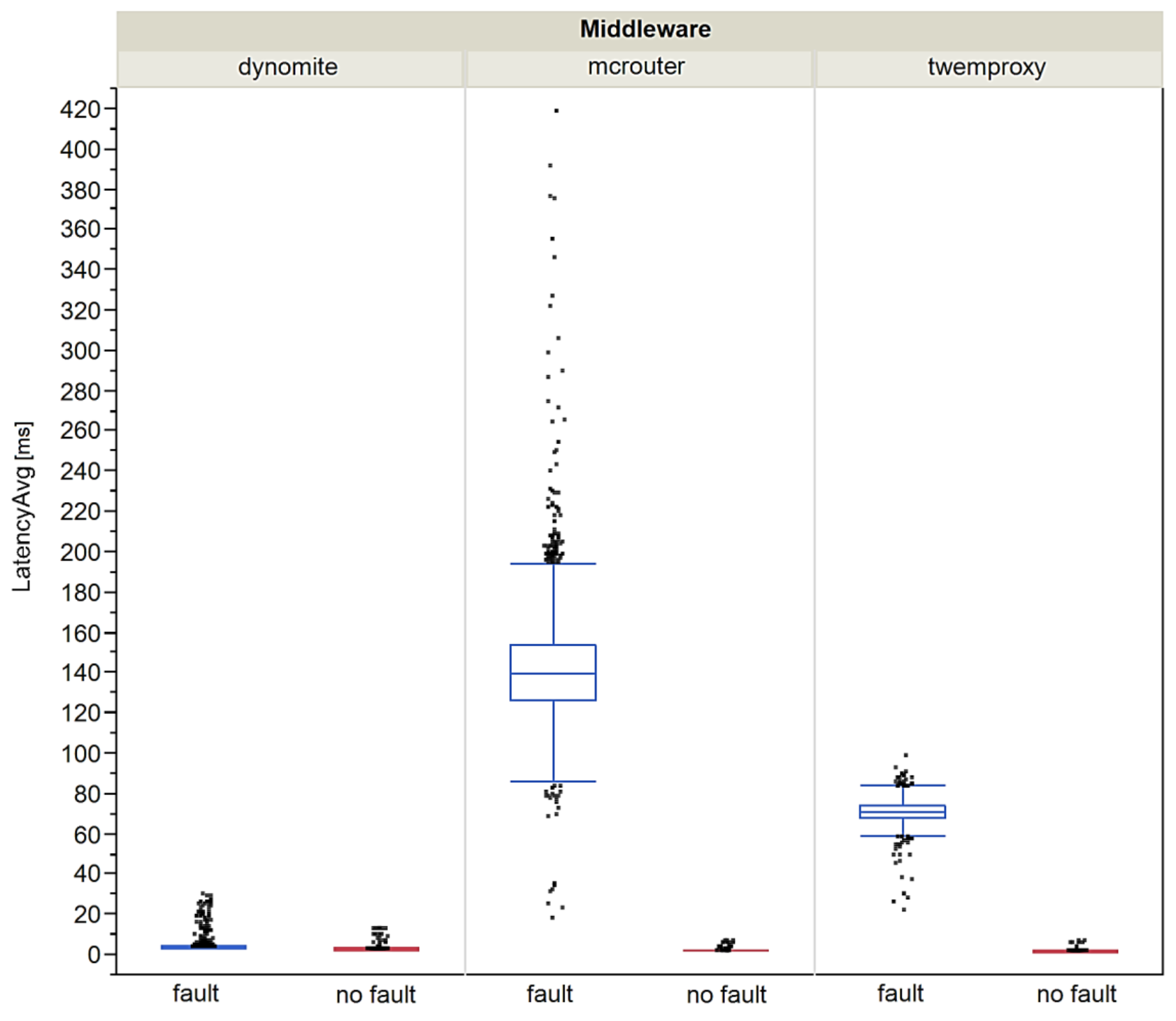}
    \caption{Performance under network link bottlenecks.}
    \label{fig:link_failure_results}
\end{figure}

In Twemproxy, bottlenecks in a subset of connections affect the caching tier as a whole system, as the overall throughput and latency degrade, showed in \figurename{}~\ref{fig:link_failure_results}. The standard deviation of latency becomes an order of magnitude higher than the average, which indicates the unbalance of the latency for different requests. Since the latency does not increase for all requests, there are not enough timeouts to trigger the ejection mechanism. The throughput degrades to $20,000$ reqs/sec, which represents a reduction of 82\% when compared to $115,000$ reqs/sec reached under a normal condition. As in the case of unbalanced overloads, in Twemproxy all nodes are impacted by the effects of the fault.

Network bottlenecks are the only kind of fault that have an impact on the performance of Mcrouter. We have a sharp increase of latency, as shown in \figurename{}~\ref{fig:link_failure_results}, and a decrease of the throughput, with some requests that experience errors. The throughput decreases from $45,000$ reqs/sec under normal conditions to $8,500$ under faults, and the average latency increases from $2.8$ ms to $142.3$ ms under faults.

Dynomite is able to manage the reduction of network bandwidth, as these faults are transparent to the clients. We found that there is no variation in throughput, and we do not have any request with errors. Comparing latency with vs. without faults (\figurename{}~\ref{fig:link_failure_results}), we notice a small increase of latency (from $2.8$ ms to $4.2$ ms under faults) which cannot be considered statistically significant. This behavior is due to the peer-to-peer architecture adopted by Dynomite. Indeed, one of the main advantages of the peer-to-peer architecture is that it increases tolerance to faults related to a specific connection. By creating several connections between nodes in a rack, the bandwidth reduction of few connections does not affect performance, since requests are forwarded to the other connections. Thus, this approach is effective at isolating the problem. The effects of the fault do not propagate across the system, which is why the performance of the caching tier as a whole is not significantly affected.

%TWEMPROXY
%- Link bandwidth bottleneck ("some connections" -> 4?)
%---> Ejection mechanism is not triggered
%---> Significant throughput reduction

%MCROUTER
%- Link bandwidth bottleneck
%---> Cannot reliably detect slow nodes, latency degrades

%DYNOMITE
%- Link bandwidth bottleneck
%---> Slow nodes have negligible impact, peer-to-peer topology isolates the problem

\section{Related work}
\label{sec:related}
Over the last few years, NoSQL data stores have found widespread adoption \cite{ davoudian2018survey }. In contrast to traditional databases, these storage systems typically sacrifice consistency in favor of latency and availability \cite {domaschka2014reliability } as in the CAP theorem \cite{ bermbach2011eventual }, so that they only guarantee eventual consistency and promise high scalability and low cost. 

Modern applications are increasingly dependent on critical data stored with these technologies. For example, these solutions allow Netflix to scale their production systems up to 50 Cassandra clusters with more than 500 nodes, and a distributed caching tier, composed by tens of thousands of Memcached instances serving about 30 million requests per second \cite{netflix2011nosql,netflix2014dynomite,netflix2016caching}. To deliver high availability and performance, many replication schemes are used to tune different levels of consistency and fault-tolerance \cite{ esteves2012quality }. In this context, companies such as Facebook, Twitter and Netflix develop and use middleware platforms \cite{ nishtala2013scaling, rajashekharcaching, papapanagiotou2018ndbench }, such as the three we compared in this work, to ease the adoption of key-value datastores at large scale. 

Performance of distributed datastores are often studied in terms of large-scale either by real deployment such as at Facebook \cite{ shankar2015benchmarking} or by simulation \cite{ wang2013consistency, wang2013using}. Klein et al. \cite{klein2015performance} compare the performance and the scalability of three popular distributed datastores (i.e., MongoDB, Cassandra and Riak) to study how the tunable consistency models affect the solution performance. Studies on real workloads pointed out the problem of unbalanced load in distributed datastore nodes \cite{ atikoglu2012workload, cooper2010benchmarking }. Many studies focus on the load balancing optimizations to prevent this issue. \emph{SPORE} \cite{hong2013understanding} is a solution to hot-spot problems due to a highly skewed workload. \emph{SPORE} modifies the traditional Memcached behavior implementing advanced data replication strategies based on key popularity. \emph{Zhang et al.} \cite{zhang2014load} propose a solution to load imbalance due to a hot-spot workload and server heterogeneity. \emph{NetKV} \cite{zhang2016netkv}, is an accelerated proxy to inspect key requests and analyze the datastore workload to replicate hot-spot keys on multiple servers, in order to limit the load unbalancing due to the workload skewness. \emph{MBal} \cite{cheng2015memory} is a novel in-memory datastore architecture aiming to resolve load unbalancing problems within the datastore tier itself. This architecture includes a centralized coordinator that monitor the system state and applies data replication and migration strategies among not only the distributed instances but also at thread level among the CPU cores within each node. More recent solutions take advantage of emerging RDMA hardware technology as an opportunity to design novel load balancing algorithms \cite{ kalia2014using, ghandeharizadeh2019scaling } to mitigate the effect of load unbalancing.

Testing the robustness of large-scale distributed systems is a challenging issue, since it is difficult to simulate large deployments with tests. Recently, many companies adopted fault injection in production systems to uncover architectural problems that could not be identified by testing in a controlled environment \cite{ basiri2016chaos, reinhold2016rewriting, sverdlik2014facebook }. Chaos Monkey \cite{ bennett2015netflix } and Simoorg \cite{ linkedln2016smoorg } are examples of fault injectors developed respectively by Netflix and LinkedIn to perform these kind of tests. Microsoft also provides a fault injector as a service \cite{ microsoft2017faultanalysis } to test applications running on their Azure IaaS cloud platform.

Fault-tolerance of distributed datastores is recurring topic in the scientific literature \cite{ beyer2011testing, nelubin2013nosql }. Ventura and Antunes \cite{ ventura2016experimental } adopted fault injection to assess the dependability of three widely-used NoSQL datastores (MongoDB, Cassandra, and Redis), and demonstrated that simple faults, such as the restart of an instance of the datastore, could affect the integrity of the whole system. Our work focuses on how faults impact on the middleware layer. Moreover, in this work we consider not only faults in the datastore system, such as the crash of Memcached servers, but also performance problems, such as unbalanced overloads and network bandwidth bottlenecks, which are both difficult to detect and very common in large deployments \cite{ huang2017gray }.

\section{Conclusion}
\label{sec:conclusion}
% !TEX root = ../main.tex
% !TEX encoding = UTF-8 Unicode

The criticality of large-scale distributed systems, where components fail routinely, mandates the use of technologies able to transparently manage failures and assure high performance. In this paper, we investigated fault-tolerance and performance concerns in \emph{middleware} for distributed caches, which has recently been emerging as a solution to deploy in-memory caches at a large scale. We conducted fault injection experiments on three middleware platform from major service providers, by simulating both crashes and performance faults. 
The experimental results lead us to the following lessons learned:

\begin{itemize}
    \item \textbf{Differences in performance and availability trade-offs}. Even if all of the middleware platforms provide fault-tolerance, fault injection revealed that they exhibit different behaviors under faults. These differences are implicitly due to their different architectures and mechanisms, but their implications are not self-evident, and knowing about them is crucial for designers to achieve performance and availability goals. We found that Twemproxy and Dynomite can escalate failures of the nodes into failures of the caching tier, either as cache misses or as service errors; instead, Mcrouter enforces full transparency of the faults, but at the cost of a worse latency under some of the fault scenarios.
    
    \item \textbf{Criticality of performance faults}. The middleware platforms are most effective at addressing crashes, but that they are more vulnerable against performance faults, such as unbalanced overloads and network bottlenecks. These faults are not systematically managed by the middleware, as this is a cross-cutting concern between the middleware and the OS, e.g., they are not explicitly notified by the OS, but can only be detected by monitoring resource utilization. These faults even caused cascading effects across the caching tier, turning into performance degradation experienced by the clients.
    
    \item \textbf{Importance of testing for performance issues at a large scale} Beyond the issues mentioned above at the architectural and fault-management levels, the middleware platforms can also suffer from implementation bugs that only become apparent when testing them at a large scale. In our tests, we found that Twemproxy exhibits a significant performance degradation under unbalanced overloads, and that the performance degradation is not simply due to the reduction of resources in the system, but it is exacerbated by a bottleneck in the software. Thus, it is important to not only rely on architectural reviews, but also to actually test the platforms at a large scale, either in a controlled or in a production environment.% excessive performance degradation with unbalanced overloads in Twemproxy
\end{itemize}

%By testing three different middleware platforms, we were able to perform a first evaluation of the techniques used for fault management and fault tolerance. It was interesting to highlight that one of the most used techniques to increase availability, data replication, does not increase the latency, which is one of the parameters that most affects the QoE. Furthermore, by varying the scale of the experiment we have been able to observe unexpected effects in some cases. For example, using Dynomite when one of the Memcached nodes crash causes errors that on small scale do not exist. This aspect highlights the importance of accurate testing even in fault scenarios, otherwise, problems like this will be discovered in production. The injection of different types of faults has allowed us to highlight how the different architectures and design choices lead to a different behavior in the same failure scenario.

%For instance, Mcrouter is more robust concerning the node crash or overload but because of its client-server architecture, it does not tolerate bandwidth reductions. Whereas, thanks to the peer-to-peer architecture, this type of fault is well tolerated by Dynomite.

%\section*{Acknowledgments}
%Omitted for double-blind review.

\bibliographystyle{IEEEtran}
\bibliography{bibliography}

\begin{thebibliography}{10}
\providecommand{\url}[1]{#1}
\csname url@samestyle\endcsname
\providecommand{\newblock}{\relax}
\providecommand{\bibinfo}[2]{#2}
\providecommand{\BIBentrySTDinterwordspacing}{\spaceskip=0pt\relax}
\providecommand{\BIBentryALTinterwordstretchfactor}{4}
\providecommand{\BIBentryALTinterwordspacing}{\spaceskip=\fontdimen2\font plus
\BIBentryALTinterwordstretchfactor\fontdimen3\font minus
  \fontdimen4\font\relax}
\providecommand{\BIBforeignlanguage}[2]{{%
\expandafter\ifx\csname l@#1\endcsname\relax
\typeout{** WARNING: IEEEtran.bst: No hyphenation pattern has been}%
\typeout{** loaded for the language `#1'. Using the pattern for}%
\typeout{** the default language instead.}%
\else
\language=\csname l@#1\endcsname
\fi
#2}}
\providecommand{\BIBdecl}{\relax}
\BIBdecl

\bibitem{souders2019velocity}
\BIBentryALTinterwordspacing
S.~Souders, ``Velocity and the bottom line,'' 2009, accessed on: 2019-11-11.
  [Online]. Available:
  \url{http://radar.oreilly.com/2009/07/velocity-making-your-site-fast.html}
\BIBentrySTDinterwordspacing

\bibitem{fitzpatrick2004distributed}
B.~Fitzpatrick, ``Distributed caching with memcached,'' \emph{Linux journal},
  vol. 2004, no. 124, p.~5, 2004.

\bibitem{anwar2020customizable}
A.~Anwar, Y.~Cheng, H.~Huang, J.~Han, H.~Sim, D.~Lee, F.~Douglis, and A.~R.
  Butt, ``Customizable scale-out key-value stores,'' \emph{IEEE Transactions on
  Parallel and Distributed Systems}, vol.~31, no.~9, 2020.

\bibitem{twemproxyrepo}
\BIBentryALTinterwordspacing
{Twitter/Twemproxy on Github}. [Online]. Available:
  \url{https://github.com/twitter/twemproxy}
\BIBentrySTDinterwordspacing

\bibitem{mcrouterrepo}
\BIBentryALTinterwordspacing
{Facebook/Mcrouter on Github}. [Online]. Available:
  \url{https://github.com/facebook/mcrouter}
\BIBentrySTDinterwordspacing

\bibitem{dynomiterepo}
\BIBentryALTinterwordspacing
{Netflix/Dynomite on Github}. [Online]. Available:
  \url{https://github.com/Netflix/dynomite}
\BIBentrySTDinterwordspacing

\bibitem{beyer2011testing}
F.~Beyer, A.~Koschel, C.~Schulz, M.~Sch{\"a}fer, I.~Astrova, A.~Reich
  \emph{et~al.}, ``Testing the suitability of cassandra for cloud computing
  environments consistency, availability and partition tolerance,'' 2011.

\bibitem{nelubin2013nosql}
D.~Nelubin and B.~Engber, ``Nosql failover characteristics: Aerospike,
  cassandra, couchbase, mongodb,'' \emph{Thumbtack Technology}, 2013.

\bibitem{ventura2016experimental}
L.~Ventura and N.~Antunes, ``Experimental assessment of nosql databases
  dependability,'' in \emph{2016 12th European Dependable Computing Conference
  (EDCC)}.\hskip 1em plus 0.5em minus 0.4em\relax IEEE, 2016, pp. 161--168.

\bibitem{huang2017gray}
P.~Huang, C.~Guo, L.~Zhou, J.~R. Lorch, Y.~Dang, M.~Chintalapati, and R.~Yao,
  ``Gray failure: The achilles' heel of cloud-scale systems,'' in
  \emph{Proceedings of the 16th Workshop on Hot Topics in Operating Systems},
  2017, pp. 150--155.

\bibitem{gunawi2018fail}
H.~S. Gunawi, R.~O. Suminto, R.~Sears, C.~Golliher, S.~Sundararaman, X.~Lin,
  T.~Emami, W.~Sheng, N.~Bidokhti, C.~McCaffrey \emph{et~al.}, ``Fail-slow at
  scale: Evidence of hardware performance faults in large production systems,''
  \emph{ACM Transactions on Storage (TOS)}, vol.~14, no.~3, pp. 1--26, 2018.

\bibitem{chi2017hashing}
L.~Chi and X.~Zhu, ``Hashing techniques: A survey and taxonomy,'' \emph{ACM
  CSUR}, 2017.

\bibitem{tanenbaum2007distributed}
A.~S. Tanenbaum and M.~Van~Steen, \emph{Distributed systems: principles and
  paradigms}.\hskip 1em plus 0.5em minus 0.4em\relax Prentice-Hall, 2007.

\bibitem{xu2013characterizing}
Y.~Xu, E.~Frachtenberg, S.~Jiang, and M.~Paleczny, ``Characterizing facebook's
  memcached workload,'' \emph{IEEE Internet Computing}, vol.~18, no.~2, pp.
  41--49, 2013.

\bibitem{lee2014cache}
M.-C. Lee, F.-Y. Leu, and Y.-P. Chen, ``Cache replacement algorithms for
  youtube,'' in \emph{2014 IEEE 28th International Conference on Advanced
  Information Networking and Applications}.\hskip 1em plus 0.5em minus
  0.4em\relax IEEE, 2014, pp. 743--750.

\bibitem{pinterest2013building}
\BIBentryALTinterwordspacing
J.~Carroll. (2013) {Building Pinterest in the cloud}. [Online]. Available:
  \url{https://medium.com/@Pinterest_Engineering/building-pinterest-in-the-cloud-6c7280dcc196}
\BIBentrySTDinterwordspacing

\bibitem{rajashekharcaching}
M.~Rajashekhar, ``Caching at twitter and moving towards a persistent, in-memory
  key-value store.''

\bibitem{papapanagiotou2018ndbench}
I.~Papapanagiotou and V.~Chella, ``Ndbench: Benchmarking microservices at
  scale,'' \emph{arXiv preprint arXiv:1807.10792}, 2018.

\bibitem{Bauer2012RAC}
E.~Bauer and R.~Adams, \emph{Reliability and Availability of Cloud Computing},
  1st~ed.\hskip 1em plus 0.5em minus 0.4em\relax Wiley-IEEE Press, 2012.

\bibitem{hong2013understanding}
Y.-J. Hong and M.~Thottethodi, ``Understanding and mitigating the impact of
  load imbalance in the memory caching tier,'' in \emph{SoCC}.\hskip 1em plus
  0.5em minus 0.4em\relax ACM, 2013.

\bibitem{huang2014characterizing}
Q.~Huang, H.~Gudmundsdottir, Y.~Vigfusson, D.~A. Freedman, K.~Birman, and
  R.~van Renesse, ``Characterizing load imbalance in real-world networked
  caches,'' in \emph{Proceedings of the 13th ACM Workshop on Hot Topics in
  Networks}.\hskip 1em plus 0.5em minus 0.4em\relax ACM, 2014, p.~8.

\bibitem{oonhawat2017hotspot}
B.~Oonhawat and N.~Nupairoj, ``Hotspot management strategy for real-time log
  data in {MongoDB},'' in \emph{ICACT}.\hskip 1em plus 0.5em minus 0.4em\relax
  IEEE, 2017.

\bibitem{cotroneo2019overload}
D.~Cotroneo, R.~Natella, and S.~Rosiello, ``Overload control for virtual
  network functions under cpu contention,'' \emph{Future Generation Computer
  Systems}, vol.~99, 2019.

\bibitem{cloudsuiterepo}
\BIBentryALTinterwordspacing
{CloudSuite Data Benchmark at Github}. [Online]. Available:
  \url{https://github.com/parsa-epfl/cloudsuite/tree/master/datasets/twitter-dataset-graph}
\BIBentrySTDinterwordspacing

\bibitem{memaslap}
\BIBentryALTinterwordspacing
{Memaslap Documentation}. [Online]. Available:
  \url{http://docs.libmemcached.org/bin/memaslap.html}
\BIBentrySTDinterwordspacing

\bibitem{twitterdataset}
\BIBentryALTinterwordspacing
{Twitter memcache dataset from CloudSuite Data Benchmark}. [Online]. Available:
  \url{https://github.com/parsa-epfl/cloudsuite/tree/master/datasets/twitter-dataset-graph}
\BIBentrySTDinterwordspacing

\bibitem{dynomiteConsistency}
\BIBentryALTinterwordspacing
{Dynomite tunable consistency}. [Online]. Available:
  \url{https://github.com/Netflix/dynomite/wiki/Consistency}
\BIBentrySTDinterwordspacing

\bibitem{davoudian2018survey}
A.~Davoudian, L.~Chen, and M.~Liu, ``A survey on nosql stores,'' \emph{ACM
  Computing Surveys (CSUR)}, vol.~51, no.~2, pp. 1--43, 2018.

\bibitem{domaschka2014reliability}
J.~Domaschka, C.~B. Hauser, and B.~Erb, ``Reliability and availability
  properties of distributed database systems,'' in \emph{2014 IEEE 18th
  International Enterprise Distributed Object Computing Conference}.\hskip 1em
  plus 0.5em minus 0.4em\relax IEEE, 2014, pp. 226--233.

\bibitem{bermbach2011eventual}
D.~Bermbach and S.~Tai, ``Eventual consistency: how soon is eventual?''
  \emph{Proc. 2011 MW4SOC}, 2011.

\bibitem{netflix2011nosql}
\BIBentryALTinterwordspacing
Y.~Izrailevsky. (2011) {NoSQL at Netflix}. [Online]. Available:
  \url{https://medium.com/netflix-techblog/nosql-at-netflix-e937b660b4c}
\BIBentrySTDinterwordspacing

\bibitem{netflix2014dynomite}
\BIBentryALTinterwordspacing
M.~Do, P.~Oberai, M.~Daxini, and C.~Kalantzis. (2014) {Introducing Dynomite:
  Making Non-Distributed Databases, Distributed}. [Online]. Available:
  \url{https://medium.com/netflix-techblog/introducing-dynomite-making-non-distributed-databases\
  -distributed-c7bce3d89404}
\BIBentrySTDinterwordspacing

\bibitem{netflix2016caching}
\BIBentryALTinterwordspacing
S.~Madappa, V.~Nguyen, S.~Mansfield, S.~Enugula, A.~Pratt, and F.~Siddiqi.
  (2016) {Caching for a Global Netflix}. [Online]. Available:
  \url{https://medium.com/netflix-techblog/caching-for-a-global-netflix-7bcc457012f1}
\BIBentrySTDinterwordspacing

\bibitem{esteves2012quality}
S.~Esteves, J.~Silva, and L.~Veiga, ``Quality-of-service for consistency of
  data geo-replication in cloud computing,'' in \emph{European Conference on
  Parallel Processing}.\hskip 1em plus 0.5em minus 0.4em\relax Springer, 2012,
  pp. 285--297.

\bibitem{nishtala2013scaling}
R.~Nishtala, H.~Fugal, S.~Grimm, M.~Kwiatkowski, H.~Lee, H.~C. Li, R.~McElroy,
  M.~Paleczny, D.~Peek, P.~Saab \emph{et~al.}, ``Scaling memcache at
  facebook,'' in \emph{Presented as part of the 10th $\{$USENIX$\}$ Symposium
  on Networked Systems Design and Implementation ($\{$NSDI$\}$ 13)}, 2013, pp.
  385--398.

\bibitem{shankar2015benchmarking}
D.~Shankar, X.~Lu, M.~Wasi-ur Rahman, N.~Islam, and D.~K. Panda, ``Benchmarking
  key-value stores on high-performance storage and interconnects for web-scale
  workloads,'' in \emph{2015 IEEE International Conference on Big Data (Big
  Data)}.\hskip 1em plus 0.5em minus 0.4em\relax IEEE, 2015, pp. 539--544.

\bibitem{wang2013consistency}
X.~Wang, H.~Sun, T.~Deng, and J.~Huai, ``Consistency or latency? a quantitative
  analysis of replication systems based on replicated state machines,'' in
  \emph{2013 43rd Annual IEEE/IFIP International Conference on Dependable
  Systems and Networks (DSN)}.\hskip 1em plus 0.5em minus 0.4em\relax IEEE,
  2013, pp. 1--12.

\bibitem{wang2013using}
K.~Wang, A.~Kulkarni, M.~Lang, D.~Arnold, and I.~Raicu, ``Using simulation to
  explore distributed key-value stores for extreme-scale system services,'' in
  \emph{SC'13: Proceedings of the International Conference on High Performance
  Computing, Networking, Storage and Analysis}.\hskip 1em plus 0.5em minus
  0.4em\relax IEEE, 2013, pp. 1--12.

\bibitem{klein2015performance}
J.~Klein, I.~Gorton, N.~Ernst, P.~Donohoe, K.~Pham, and C.~Matser,
  ``Performance evaluation of nosql databases: a case study,'' in
  \emph{Proceedings of the 1st Workshop on Performance Analysis of Big Data
  Systems}, 2015, pp. 5--10.

\bibitem{atikoglu2012workload}
B.~Atikoglu, Y.~Xu, E.~Frachtenberg, S.~Jiang, and M.~Paleczny, ``Workload
  analysis of a large-scale key-value store,'' in \emph{Proceedings of the 12th
  ACM SIGMETRICS/PERFORMANCE joint international conference on Measurement and
  Modeling of Computer Systems}, 2012, pp. 53--64.

\bibitem{cooper2010benchmarking}
B.~F. Cooper, A.~Silberstein, E.~Tam, R.~Ramakrishnan, and R.~Sears,
  ``Benchmarking cloud serving systems with ycsb,'' in \emph{Proceedings of the
  1st ACM symposium on Cloud computing}, 2010, pp. 143--154.

\bibitem{zhang2014load}
W.~Zhang, J.~Hwang, T.~Wood, K.~Ramakrishnan, and H.~H. Huang, ``Load balancing
  of heterogeneous workloads in memcached clusters.'' in \emph{Feedback
  Computing}, 2014.

\bibitem{zhang2016netkv}
W.~Zhang, T.~Wood, and J.~Hwang, ``Netkv: Scalable, self-managing, load
  balancing as a network function,'' in \emph{Autonomic Computing (ICAC), 2016
  IEEE International Conference on}.\hskip 1em plus 0.5em minus 0.4em\relax
  IEEE, 2016, pp. 5--14.

\bibitem{cheng2015memory}
Y.~Cheng, A.~Gupta, and A.~R. Butt, ``An in-memory object caching framework
  with adaptive load balancing,'' in \emph{Proceedings of the Tenth European
  Conference on Computer Systems}.\hskip 1em plus 0.5em minus 0.4em\relax ACM,
  2015, p.~4.

\bibitem{kalia2014using}
A.~Kalia, M.~Kaminsky, and D.~G. Andersen, ``Using rdma efficiently for
  key-value services,'' in \emph{Proceedings of the 2014 ACM conference on
  SIGCOMM}, 2014, pp. 295--306.

\bibitem{ghandeharizadeh2019scaling}
S.~Ghandeharizadeh and H.~Huang, ``Scaling data stores with skewed data access:
  Solutions and opportunities,'' in \emph{8th Workshop on Scalable Cloud Data
  Management, co-located with IEEE BigData}, 2019.

\bibitem{basiri2016chaos}
A.~Basiri, N.~Behnam, R.~De~Rooij, L.~Hochstein, L.~Kosewski, J.~Reynolds, and
  C.~Rosenthal, ``Chaos engineering,'' \emph{IEEE Software}, vol.~33, no.~3,
  pp. 35--41, 2016.

\bibitem{reinhold2016rewriting}
E.~Reinhold, ``Rewriting uber engineering: the opportunities microservices
  provide. uber engineering,'' 2016.

\bibitem{sverdlik2014facebook}
Y.~Sverdlik, ``Facebook turned off entire data center to test resiliency,''
  \emph{Data Center Knowledge}, vol.~15, 2014.

\bibitem{bennett2015netflix}
C.~Bennett and A.~Tseitlin, ``Netflix http://techblog. netflix.
  com/2012/07/chaos-monkey-released-into-wild. html (2012),'' 2015.

\bibitem{linkedln2016smoorg}
\BIBentryALTinterwordspacing
A.~Shenoy. (2016) {A Deep Dive into Simoorg: Our Open Source Failure Induction
  Framework}. [Online]. Available:
  \url{https://engineering.linkedin.com/blog/2016/03/deep-dive-Simoorg-open-source-failure-induction-framework}
\BIBentrySTDinterwordspacing

\bibitem{microsoft2017faultanalysis}
\BIBentryALTinterwordspacing
Microsoft. (2017) {Introduction to the Fault Analysis Service}. [Online].
  Available:
  \url{https://azure.microsoft.com/en-us/documentation/articles/service-fabric-testability-overview/}
\BIBentrySTDinterwordspacing

\end{thebibliography}

\end{document}